\documentclass[Physsubmission, Phys]{SciPost}

\hypersetup{
    colorlinks,
    linkcolor={red!50!black},
    citecolor={blue!50!black},
    urlcolor={blue!80!black}
}

\usepackage[bitstream-charter]{mathdesign}
\DeclareSymbolFont{usualmathcal}{OMS}{cmsy}{m}{n}
\DeclareSymbolFontAlphabet{\mathcal}{usualmathcal}

\begin{document}

\begin{center}{\Large \textbf{
Study of proton parton distribution functions at high x\\
}}\end{center}

\begin{center}
Ritu Aggarwal\textsuperscript{1$\star$}
\end{center}

\begin{center}
{\bf 1} Savitribai Phule Pune University, India
\\
* ritu.aggarwal1@gmail.com
\\
(On behalf of the ZEUS Collaboration)
\end{center}

\begin{center}
\today
\end{center}


\definecolor{palegray}{gray}{0.95}
\begin{center}
\colorbox{palegray}{
  \begin{tabular}{rr}
  \begin{minipage}{0.1\textwidth}
    \includegraphics[width=22mm]{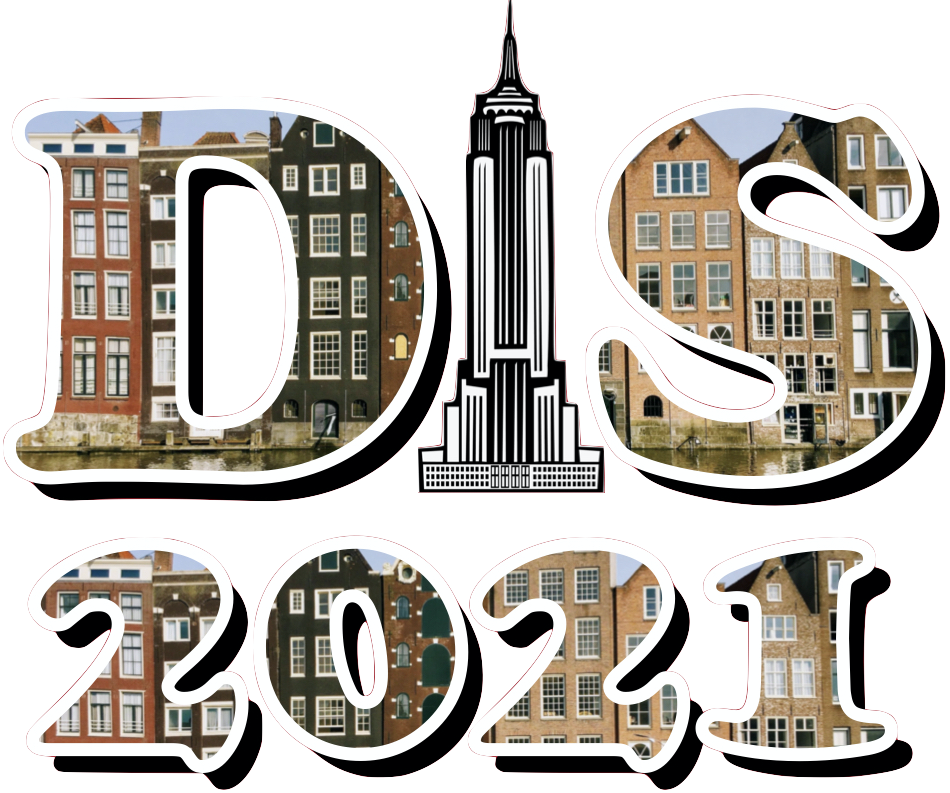}
  \end{minipage}
  &
  \begin{minipage}{0.75\textwidth}
    \begin{center}
    {\it Proceedings for the XXVIII International Workshop\\ on Deep-Inelastic Scattering and
Related Subjects,}\\
    {\it Stony Brook University, New York, USA, 12-16 April 2021} \\
    \doi{10.21468/SciPostPhysProc.?}\\
    \end{center}
  \end{minipage}
\end{tabular}
}
\end{center}

\section*{Abstract}
{\bf
At large values of $x$ the parton distribution functions (PDFs) of the proton are poorly constrained and there are considerable variations between different global fits. Data at such high $x$ have already been published by the ZEUS Collaboration, but not yet used in PDF extractions. A technique for comparing predictions based on different PDF sets to the observed number of events in the ZEUS data is presented. It is applied to compare predictions from the most commonly used PDFs to published ZEUS data at high Bjorken $x$. A wide variation is found in the ability of the PDFs to predict the observed results. A scheme for including the ZEUS high $x$ data in future PDF extractions is discussed.
}

\vspace{10pt}
\noindent\rule{\textwidth}{1pt}
\tableofcontents\thispagestyle{fancy}
\noindent\rule{\textwidth}{1pt}
\vspace{10pt}

\section{Introduction}
\label{sec:intro}

A precise information on the quark and Gluon content of proton is very important to analyze the data collected from the high energy particle collide rs that are already in operation and also those which are proposed in the future. The deep inelastic scattering (DIS) data~\cite{ref:herapdf2.0,ref:highQ2Z} collected from the HERA collide and the data collected through the fixed target experiments~\cite{ref:FixedTarget:1,ref:FixedTarget:2,ref:FixedTarget:3,ref:FixedTarget:4,ref:FixedTarget:5} have helped to understand the parton distribution functions (PDFs) inside proton. But the picture of distribution of par tons at high Bjorken $x$ is still not clear, where different theoretical PDFs  tend to show a large PDF uncertainty. Figure 1 shows the ratio of ep NC DIS events generated using commonly used modern PDFs(~\cite{ref:MMHT2,ref:CT2,ref:ABM2,ref:NNPDF2}) to HERAPDF2.0~\cite{ref:herapdf2.0} in bins of $x$ and $Q^2$ (four-momentum transferred from electron to proton). The large uncertainties at the high $x$ region as calculated from HERAPDF2.0 and  NNPDF3.1~\cite{ref:NNPDF2}, as shown in Figure 1 in the grey and red cross hatched bands respectively, depict a need of taking into account any information available in this region. The high-x $ep$ NC ZEUS data~\cite{ref:highxpaper} is one such data that can impact our understanding of structure of proton in this region.

\begin{figure}[h]
\centering
\includegraphics[width=0.65\textwidth]{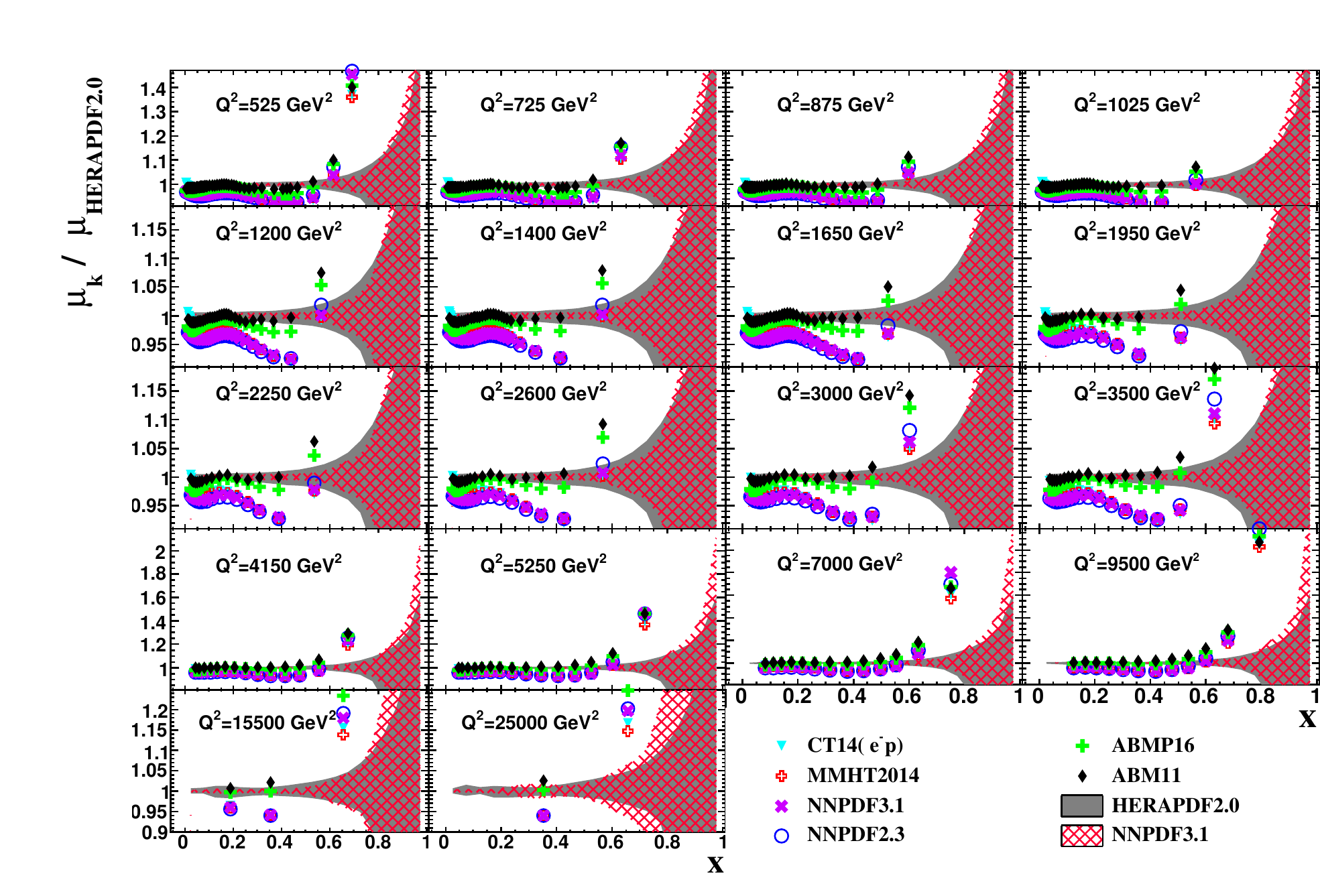}
\caption{Ratio of $e^-p$ NC DIS events generated from different PDFs to HERAPDF2.0. The PDF uncertainty from HERAPDF2.0 and NNPDF3.1 are shown in the grey band and  red cross hatched band respectively. }
\label{fig1}
\end{figure}

\section{ZEUS high $x$ NC ep data}
The ZEUS high $x$ NC data~\cite{ref:highxpaper} is a unique data as it spans the Bjorken $x$ region up to a value of 1.    In a NC ep collision event, electron and jets of final state hadrons are observed in the detector as shown in Figure 2(left). However in the case of very high $x$ events, only electron is collected in the detector as shown in Figure 2(right). These high $x$ are essentially zero jet events as the hadronic activity is very close to the beam pipe in the forward direction.

\begin{figure}[h]
\centering
\includegraphics[width=0.25\textwidth]{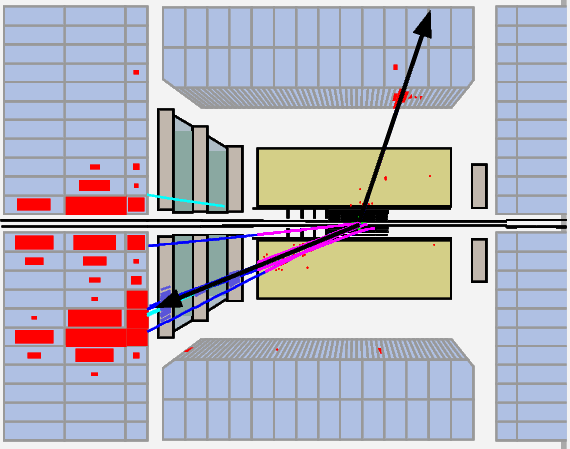}
\includegraphics[width=0.25\textwidth]{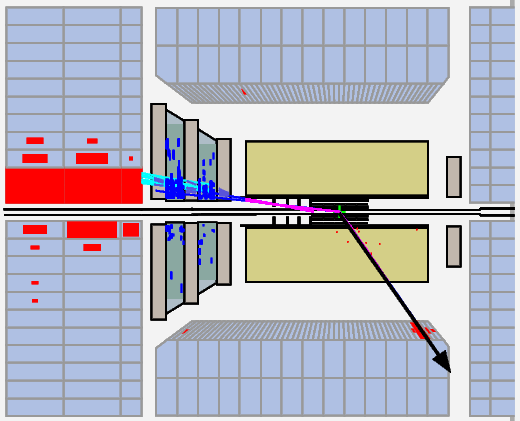}
\caption{(Left) Event display for ep NC DIS event captured by the ZEUS detector (Right) Event display for a high $x$ ep NC DIS event with no jets.}
\label{ref}
\end{figure}
These events lie in the kinematic range above the $x_{edge}$ up to 1, and for these events integrated cross section is reported. The comparison of NC $e^+p$ cross sections reported in~\cite{ref:highxpaper} to the HERAPDF1.5~\cite{ref:herapdf1.5} predictions are shown in Figure 3. Data points in blue are the double differential cross sections and in black triangles are the integrated cross sections reported for the zero jet events.

\begin{figure}[h]
\centering
\includegraphics[width=0.5\textwidth]{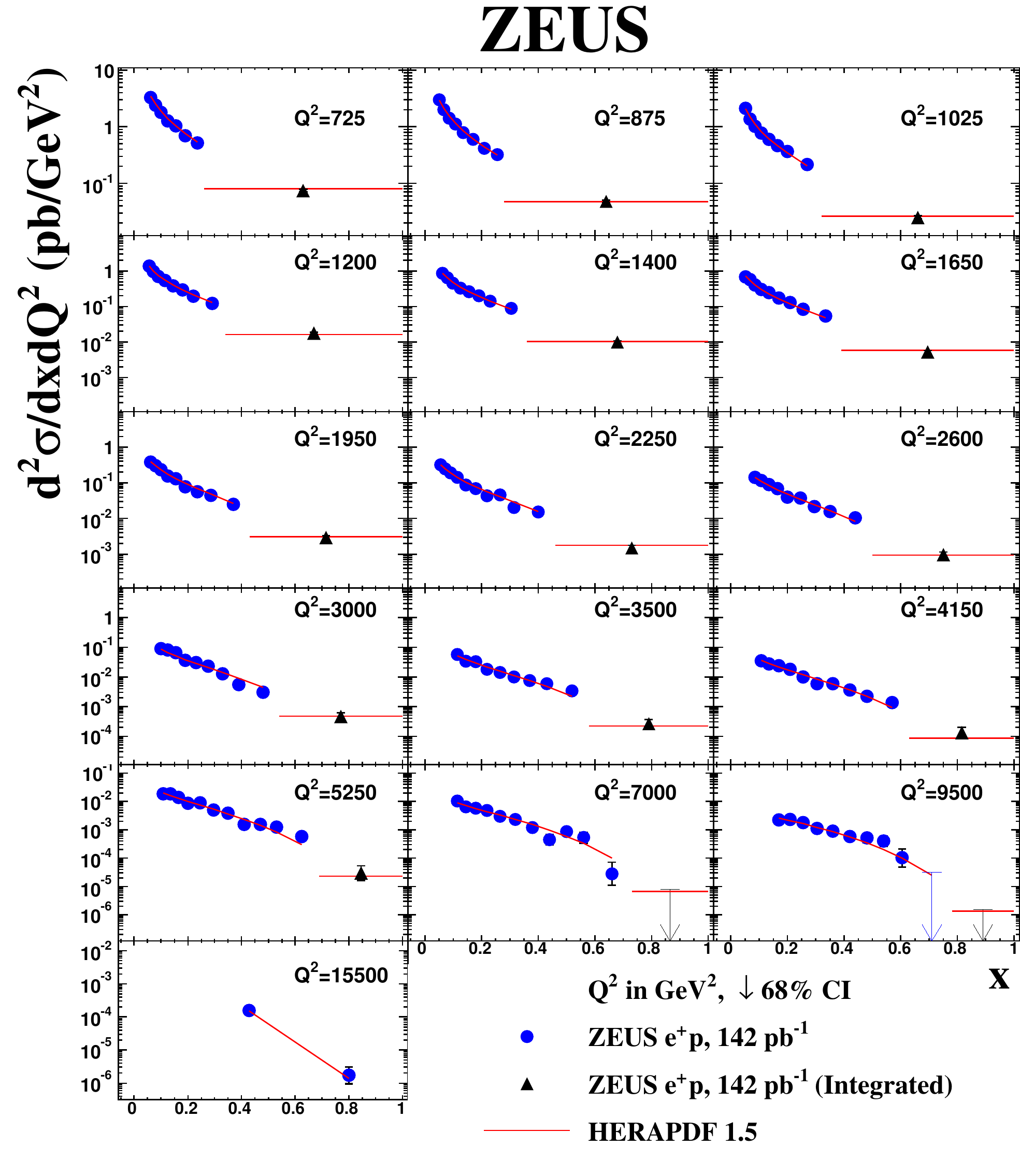}
\caption{Double differential and integrated cross section for the ZEUS $e^+p$ NC data set~\cite{ref:highxpaper}, shown in the bins of $Q^2$ and $x$. Prediction from HERAPDF1.5 are shown in the red line.}
\label{ref}
\end{figure}

\section{Transfer Matrix for ZEUS high $x$  data}
\label{sec:another}
The expected and observed number of events in the high $x$ kinematic region are very small and Poisson errors are quoted for the statistical uncertainty. To facilitate their inclusion in the PDF determination, transfer matrix~\cite{ref:highxpaper2} has been calculated for the high $x$ ZEUS data. 
If $\lambda_{l,k}$ are the Born cross section in the $l^{th}$ true $x-Q^2$ phase space interval from any theoretical PDF $k$, then the prediction for the number of events in the $j^{th}$ bin at the detector level is given as

\begin{equation}
\nu_{j,k}  \approx \sum_{l} A_{jl}\lambda_{l,k} 
\end{equation} 

Where matrix A can be divided into two parts:
\begin{equation}
A =T R
\end{equation}
Here T is called as the Transfer Matrix and R contains Radiative corrections. 
 
\subsection{Transfer Matrix}
Each element in the Transfer Matrix, T$_{ji}$ gives a probability that an event reconstructed in the $(x-Q^2)_j^{th}$ interval was generated in the true $(x-Q^2)_i^{th}$ interval and is calculated from the simulated data using the equation given below:

\begin{equation}
T_{ji} = \frac{\sum^{M_{i}}_{m=1} \omega_{m} I (m \in j)}{\sum^{M_{i}}_{m=1} \omega_{m}^{\rm {MC}}}  \; .
\label{eqn:a_ij}		
\end{equation}
 where $\omega_{m}^{\rm {MC}}$ is the weight carried by each generated event, $\omega_{m}$ includes in addition the weights due to simulation efficiencies and $I (m \in j)$ is the indicator factor whose value is 1 if event $m$ belongs to  bin $j$, else it is 0.
 
\subsection{Radiative Corrections}
Diagonal matrix R contains the radiative corrections which are applied on the Born cross sections to get number of events at the generated level. These are derived using the equation given below:

\begin{equation}
 R_{ii} = \frac{\sum^{M_{i}}_{m=1}\omega_{m}^{\rm{MC}} }{\mathcal {L^{\rm MC}} \sigma_{i,{\rm CTEQ5D}}}  
 \end{equation}

 Here $M_{i}$ are the events generated in a given $i^{th}$ true $x-Q^2$ bin, $L^{\rm MC}$ is the luminosity of  simulated data and $\sigma_{i,{\rm CTEQ5D}}$ is the Born cross section in the given bin. 

\section{Comparison of different PDFs to the high $x$ data}
\label{sec:PDFs}
Predictions for number of  events in the cross section bins are calculated from different PDFs using Equation (1) and are compared to the high $x$ data. Table 1 shows the p-values, the Bayes factor for different PDFs with respect to HERAPDF2.0 and the  corresponding  $\Delta \chi^2 $ values. A wide variation is found in the ability of different PDFs to predict the high $x$ ZEUS data. Differences in p-values and the Bayes factor are also observed for the $e^-p$ and $e^+p$ data sets. HERAPDF2.0 gives a higher p-value for the $e^-p$ data set and a lower p-value for the $e^+p$ data set as compared to other PDFs.

\begin{table}
\centering
\begin{tabular}{|c|ccc|ccc|}
\hline
 &  \multicolumn{3}{c|}{$e^-p$} & \multicolumn{3}{|c|}{$e^+p$}  \\
\hline
 PDF &   $p$-value&P1/P2 & $\Delta \chi^2 $& $p$-value&P1/P2 &  $\Delta \chi^2 $\\
\hline
  HERAPDF2.0  & $ 2.8\times10^{-2} $ & $ 1.0  $ & $ 0.0 $ & $ 0.35 $  & $ 1.0  $ & $ 0.0 $\\ 
  CT14  & $ 3.2 \times 10^{-3} $ & $ 7.6 \times 10^{-3} $ & $ 9.8 $ & $ 0.82 $  & $ 5.9 \times 10^{+5} $ & $ -27 $\\ 
  MMHT2014  & $ 2.3 \times 10^{-3} $ & $ 2.1 \times 10^{-3} $ & $ 12 $ & $ 0.82 $  & $ 4.7 \times 10^{+5} $ & $ -26 $\\ 
  NNPDF3.1  & $ 3.9 \times 10^{-4} $ & $ 3.2 \times 10^{-6} $ & $ 25 $ & $ 0.73 $  & $ 9.0 \times 10^{+4} $ & $ -23 $\\ 
  NNPDF2.3  & $ 1.3 \times 10^{-4} $ & $ 2.3 \times 10^{-7} $ & $ 31 $ & $ 0.70 $  & $ 4.2 \times 10^{+4} $ & $ -21 $\\ 
  ABMP16  & $ 2.6 \times 10^{-2} $ & $ 9.0 \times 10^{-1} $ & $ 0.21 $ & $ 0.64 $  & $ 6.1 \times 10^{+2} $ & $ -13 $\\ 
  ABM11  & $ 3.3 \times 10^{-2} $ & $ 7.2 \times 10^{-1} $ & $ 0.67 $ & $ 0.45 $  & $ 2.8  $ & $ -2.1 $\\

\hline
\end{tabular}
\caption{The  $p$-values, Bayes Factors (P1/P2) and $\Delta\chi^2$ from predictions using different PDF sets  when compared to the observed numbers of events. }
\label{tab:probs}
\end{table}

\section{Conclusion}
Transfer Matrix for the ZEUS high $x$ data is presented which can be used to calculate the  predictions for number of events in the cross section intervals from different PDFs. Using this technique, predictions from different commonly used PDFs are compared to the number of events reported in the high $x$ ZEUS data. Differences are observed at the generator level as illustrated in Figure 1 for the $ep$ data set and also at the detector level as shown in Table 1 for the $e^-p$ and $e^+p$ data sets. The observed differences and the large PDF uncertainty in the high $x$ region point towards a need of studying the effects of including the high $x$ data in PDF determination.
\section*{Acknowledgements}
Author acknowledges the support of Department of Science and Technology, Ministry of Science and Technology, India through the DST-INSPIRE Faculty grant.







\begin{thebibliography}{99}
\small

\bibitem{ref:herapdf2.0}
H1 and ZEUS Coll.,  H. Abramovicz et al., {\it Combinations of measurement of inclusive deep inelastic $e^{\pm}$ scattering cross sections and qcd analysis of hera data}, Eur. Phys. J.  {\bf C 75}, 580 (2015). doi:10.1140/epjc/s10052-015-3710-4.

\bibitem{ref:highQ2Z}
ZEUS Coll., H. Abramowicz et al., {\it Measurement of high-$Q^2$ neutral current deep inelastic $e^+p$ scattering cross sections with a longitudinally polarized positron beam at HERA}, Phys. Rev. {\bf D 87}, 052014 (2013). doi:10.1103/PhysRevD.87.052014.
\bibitem{ref:FixedTarget:1}
V. N. Gribov  and  L. N. Lipatov, {\it Deep inelastic ep scattering in perturbation theory}, Sov. J. Nucl. Phys.  {\bf 15}, 438 (1972). doi:10.1016/0370-2693(71)90576-4.

\bibitem{ref:FixedTarget:2}
 V. N. Gribov and L. N. Lipatov, {\it $e^+ e^-$ pair annihilation and deep inelastic e p scattering in perturbation theory}, Sov. J. Nucl. Phys.  {\bf 15}, 675 (1972).

\bibitem{ref:FixedTarget:3}
 L. N. Lipatov, {\it The parton model and perturbation theory}, Sov. J. Nucl. Phys. {\bf 20}, 94 (1975).

\bibitem{ref:FixedTarget:4}
Yu. L. Dokshitzer, {\it Calculation of the structure functions for deep inelastic scattering and $e^+e^-$ annihilation by perturbation theory in quantum chromodynamics (in russian)}, Sov. Phys. JETP  {\bf 46}, 641 (1977).

\bibitem{ref:FixedTarget:5}
G. Altarelli, and G. Parisi, {\it Asymptotic freedom in parton language}, Nucl. Phys.  {\bf B 126}, 298 (1977).

\bibitem{ref:MMHT2}
 L. Harland-Lang, A. D. Martin, P. Motylinski and R. Thorne, {\it Parton distributions in the LHC era: MMHT 2014 PDFs} {\bf 75}, 204 (2014)  doi:10.1140/epjc/s10052-015-3397-6. 

\bibitem{ref:CT2}
 M. Guzzi et al., {\it New parton distributions for collider physics}, Phys. Rev. {\bf D 82}, 074024 (2010), doi:10.1103/PhysRevD.82.074024. 

\bibitem{ref:ABM2}
 S. Alekhin et al., {\it Parton distribution functions and benchmark cross sections at NNLO} Phys. Rev. {\bf D 86}, 054009 (2012), doi:10.1103/PhysRevD.86.054009. 

\bibitem{ref:NNPDF2}
 R. D. Ball et al., NNPDF Collaboration, {\it Parton distributions for the LHC Run II} JHEP  {\bf 1004}, 040 (2015), doi:10.1007/JHEP04(2015)040. 

\bibitem{ref:highxpaper}
ZEUS Coll., H. Abramowicz et al., {\it Measurement of neutral current $e^{\pm}p$ cross sections at high Bjorken
x with the ZEUS detector} Phys. Rev.  {\bf D 89}, 072007 (2014). doi:10.1103/PhysRevD.89.072007.
\bibitem{ref:herapdf1.5}
H1 and ZEUS Coll.,  H. Abramovicz et al., {\it Combined inclusive diffractive cross sections measured with forward proton spectrometers in deep inelastic ep scattering at HERA}, Eur. Phys. J.  {\bf C 72}, 2175 (2012). doi:10.1140/epjc/s10052-012-2175-y.

\bibitem{ref:highxpaper2}
ZEUS Coll., H. Abramowicz et al., {\it study of proton parton distribution functions at high x using ZEUS data } Phys. Rev.  {\bf D 101}, 112009 (2020). doi:10.1103/PhysRevD.101.112009.

\end{thebibliography}

\nolinenumbers

\end{document}